\documentstyle [sprocl,epsf,psfig]{article}

\def\be{\begin{equation}}
\def\ee{\end{equation}}
\def\bea{\begin{eqnarray}}
\def\eea{\end{eqnarray}}
\title{Pancharatnam revisited}
\author{Erik Sj\"{o}qvist}

\address{Department of Quantum Chemistry, Uppsala University, \\  
Box 518, Se-751 20 Uppsala, Sweden}

\begin{document}
\maketitle

\abstracts{Some recent ideas concerning Pancharatnam's prescription 
of relative phase between quantal states are delineated. 
Generalisations to mixed states and entangled two-photon states 
are discussed. An experimental procedure to test the geometric 
phase as a Pancharatnam relative phase is described. We further 
put forward a spatial split-beam dual to Pancharatnam's relative 
phase.}    

\section{Introduction}
Consider a pair of vectors $|A\rangle$ and 
$|A\rangle' = e^{i\alpha}|A\rangle$ representing one and 
the same quantal state $A$. Let us ask: what is their relative 
phase? Probably everyone would agree that the answer to this 
question is $\alpha$. Less obvious, however, is the case when 
the pair of vectors represent different quantal states. 
Apparently no one had asked this question until Pancharatnam 
\cite{pancharatnam56} came up with a physical prescription for 
the relative phase between distinct polarisation states of light. 
Although his analysis concerned classical light it has been 
realised that Pancharatnam's concept of relative phase has a 
quantal counterpart with surprisingly rich structure related 
to the geometry of the quantum state space. This counterpart, 
known as quantum parallel transport, is the origin of the 
geometric phase first discovered for cyclic adiabatic evolution 
\cite{berry84} and later generalised to nonadiabatic 
\cite{aharonov87} and noncyclic \cite{samuel88} pure 
state evolutions, as well as to mixed quantal states 
\cite{uhlmann86,sjoqvist00a}.  

The purpose of this report is to describe some recent ideas 
concerning the concept of relative phase. To do this we first 
describe Pancharatnam's original idea and how it relates to 
the geometry of quantum state space. The basic observation  
in this context is the experimental fact \cite{wagh98} that 
the maximum of the interference oscillations, obtained when 
a variable phase shift is introduced to one of the interfering 
states $A$ and $B$, is shifted when $A \neq B$. This shift 
constitutes the desired concept of relative phase. To develop 
this idea in detail, let us suppose that $|A\rangle$ and 
$|B\rangle$ are two normalised Hilbert space representatives 
of the nonorthogonal states $A$ and $B$, respectively, and 
assume further that $|A\rangle$ is exposed to the $U(1)$ 
shift $e^{i\chi}$. The resulting interference pattern is 
determined by 
\begin{equation}
I = \Big| e^{i\chi}|A\rangle + |B\rangle \Big|^{2} = 
2 + 2 |\langle A|B \rangle| 
\cos [ \chi - \arg \langle A|B \rangle ]
\label{eq:pureinterfer}
\end{equation}  
that attains its maximum at the Pancharatnam relative phase 
$\phi \equiv \arg \langle A|B \rangle$. This phase reduces 
naturally to the $U(1)$ case whenever $|B\rangle = 
e^{i\alpha} |A\rangle$ as it yields $\phi = \alpha$.  

Pancharatnam's prescription of relative phase has a peculiar 
property that arises when two in-phase ($\phi =0$) vectors 
$|A\rangle$ and $|B\rangle$ are sent through a polariser 
that projects onto a third state $C$. The resulting state 
vectors are $|C\rangle' = |C\rangle \langle C|A\rangle$ and 
$|C\rangle'' = |C\rangle \langle C|B\rangle$. Their relative 
phase becomes 
\begin{equation}
\arg \, ' \! \langle C|C \rangle'' =  
\arg \langle A|C \rangle \langle C|B \rangle\langle B|A \rangle 
\equiv \Delta (A,B,C) , 
\label{eq:chain} 
\end{equation}
where we have used the assumption that $\arg \langle A|B \rangle = 0$. 
The right-hand side $\Delta (A,B,C)$ of Eq. (\ref{eq:chain}) is 
independent of the choice of Hilbert space representatives of the 
three states $(A,B,C)$ and is therefore a property of the quantum 
state space. $\Delta (A,B,C)$ further fulfils the following two 
important properties that makes it related to oriented area 
\cite{mukunda93}: 
\begin{itemize} 
\item[a)] It is additive: $\Delta (A,B,C,D) = \Delta (A,B,C) + 
\Delta (A,C,D)$. 
\item[b)] It depends upon orientation: 
$\Delta (A,C,B) = -\Delta (A,B,C)$. 
\end{itemize} 
Indeed, in the case of qubits (two-level systems) these properties 
lead to the following natural geometric interpretation of 
$\Delta (A,B,C)$ known as the geometric phase. Let us take $A$ 
at the north pole, $B$ with polar angles $( \theta , \varphi )$, 
and $C=B+dB$ with polar angles $( \theta + d\theta , \varphi + 
d\varphi )$ being infinitesimally close to $B$ on the Bloch sphere. 
It  yields $\Delta (A,B,B+dB) = -\frac{1}{2} (1-\cos \theta) 
d\varphi$, which is minus one-half the solid angle enclosed by 
the spherical triangle defined by the three points $A,B,B+dB$ 
on the Bloch sphere. Thus, for a finite spherical triangle 
defined by any $A,B,C$ on the Bloch sphere one may divide 
the geodesic line connecting $B$ and $C$ into small pieces 
and use the properties a) and b) to obtain  
\begin{equation}
\Delta (A,B,C) = - \Delta (A,C,B) = -\frac{1}{2} \Omega (A,B,C) , 
\label{eq:puresolid}
\end{equation}
where $\Omega (A,B,C)$ is the enclosed solid angle. 

In the next section we apply the above ideas to mixed quantal 
states. Section {\bf 3} contains a description of Pancharatnam's 
relative phase for nonclassical two-photon polarisation states. 
Generalising the above solid angle formula to any continuous 
path on the Bloch sphere raises the question how such a 
quantity can be tested experimentally. In section {\bf 4} 
a general procedure how this can be achieved is described. 
A spatial dual to Pancharatnam's original idea concerning 
internal states is discussed in section {\bf 5}. The paper 
ends with the conclusions.  

\section{Generalisation to mixed states}
Suppose a mixed quantal state evolves as $\rho_{A} \longrightarrow 
\rho_{B} = U\rho_{A}U^{\dagger}$ with $U$ unitary. How is 
Pancharatnam's prescription of relative phase generalised 
to such states? Here, we discuss such a generalisation that 
was discovered in \cite{sjoqvist00a} in the context of 
interferometry. 

The mixed state phase in \cite{sjoqvist00a} is based on the 
observation that the evolution governed by the unitarity 
$U$ of any density operator may be described as   
\begin{equation}
\rho_{A} = \sum_{k} w_{k} |A_{k} \rangle \langle A_{k}| 
\longrightarrow \rho_{B} = 
\sum_{k} w_{k} |B_{k} \rangle \langle B_{k}| , 
\end{equation}
where each $|B_{k}\rangle = U|A_{k} \rangle$. Evidently, 
each such orthonormal pure state component of the density 
operator contributes to the interference according to Eq. 
(\ref{eq:pureinterfer}). Thus, the total interference 
profile becomes 
\begin{eqnarray} 
I & = & \sum_{k} w_{k} \Big| e^{i\chi}|A_{k}\rangle + 
|B_{k}\rangle \Big|^{2} 
\nonumber \\ 
 & = & 2 + 2 \sum_{k} w_{k} |\langle A_{k}|B_{k} \rangle| 
\cos [ \chi - \arg \langle A_{k}|B_{k} \rangle ] , 
\end{eqnarray}  
where we have used that the $w_{k}$'s sum up to unity. 
This takes a more transparent and explicitly basis independent 
form when writing the interference profile in terms of the 
quantity ${\mbox{Tr}} ( U\rho_{A} )$ as 
\begin{equation} 
I = 2 + 2| {\mbox{Tr}} ( U\rho_{A} ) | 
\cos [\chi - \arg {\mbox{Tr}} ( U\rho_{A} )] . 
\label{eq:mixedinterfer}
\end{equation}

The important observation from Eq. (\ref{eq:mixedinterfer})
is that the interference oscillations produced by the 
variable $U(1)$ phase $\chi$ is shifted by $\Phi = 
\arg {\mbox{Tr}} ( U \rho_{A} )$ for any mixed 
state $\rho_{A}$. This shift reduces to Pancharatnam's 
original prescription when  
$\rho_{A} = |A \rangle \langle A| \longrightarrow 
\rho_{B} = |B \rangle \langle B|$. These two facts are 
the central properties for $\Phi$ being a natural 
generalisation of Pancharatnam's relative phase to 
mixed states. Moreover, the visibility of the interference 
pattern is $|{\mbox{Tr}} ( U \rho_{A} )| \geq 0$, 
which reduces to the expected $|\langle A|B \rangle|$ in 
Eq. (\ref{eq:pureinterfer}) for pure states. 

The oriented area introduced in Eq. (\ref{eq:chain}) may 
be generalised to the sequence $\rho_{A} \longrightarrow 
\rho_{B} \longrightarrow \rho_{C}$ of nondegenerate 
\footnote{If the density operators are degenerate their 
eigenbases are not unique and $\Delta (A,B,C)$ in Eq. 
(\ref{eq:mixedarea}) below becomes undefined} density 
operators, all with the same set of eigenvalues $\{ w_{k} \}$, 
but with different sets of eigenvectors $\{ |A_{k} \rangle \}$, 
$\{ |B_{k} \rangle \}$, $\{ |C_{k} \rangle \}$. First, let 
us introduce the quantities $\Delta (A_{k},B_{k},C_{k})$ 
for each set of pure state components corresponding to 
one and the same eigenvalue $w_{k}$. In terms of these 
we may take the mixed state generalisation of Eq. 
(\ref{eq:chain}) as 
\begin{equation} 
\Delta (A,B,C) = 
\arg \Big[ \sum_{k} w_{k} |\langle A_{k}|U|A_{k} \rangle| 
e^{i\Delta (A_{k},B_{k},C_{k})} \Big] , 
\label{eq:mixedarea}
\end{equation} 
where $U$ now takes $\rho_{A}$ to $\rho_{C}$ via $\rho_{B}$. Note 
that $\Delta (A,B,C)$ in Eq. (\ref{eq:mixedarea}) is nonadditive 
due to its nonlinear dependence upon the additive pure state 
quantities $\Delta (A_{k},B_{k},C_{k})$, but it depends upon 
the orientation of the unitary path $U$ as it fulfils 
$\Delta (A,C,B) = -\Delta (A,B,C)$. 

For a mixed qubit state $\rho_{A} = \frac{1}{2} (1+r) |A_{0}\rangle 
\langle A_{0}| + \frac{1}{2} (1-r) |A_{1} \rangle \langle A_{1} |$, 
where $\langle A_{k}|A_{l} \rangle = \delta_{kl}$ and 
$-1\leq r \leq 1$, we have $|\langle A_{0}|U|A_{0} \rangle| 
= |\langle A_{1}|U|A_{1} \rangle|$ and 
$\Delta (A_{0},B_{0},C_{0}) = -\Delta (A_{1},B_{1},C_{1}) = 
-\frac{1}{2} \Omega (A_{0},B_{0},C_{0}) \equiv 
-\frac{1}{2} \Omega$. Thus, if the $\rho$'s are nondegenerate 
($r\neq 0$) we obtain  
\begin{equation} 
\Delta (A,B,C) = 
-\arctan \Big[ r \tan \frac{\Omega}{2} \Big] , 
\label{eq:mixedsolid}
\end{equation} 
which is the mixed state generalisation of Eq. (\ref{eq:puresolid}) 
and reduces to the pure state case when $|r|=1$. 

\section{Two-photon relative phase}
Pancharatnam's original work applies to polarised classical 
light waves as well as to quantal single photon states. In 
this section we generalise this situation to nonclassical 
entangled two-photon polarisation states. 

Consider first a photon pair being prepared in a product 
polarisation state $|\Pi_{0}\rangle = |A\rangle \otimes 
|A'\rangle \equiv |AA'\rangle$. Suppose each of these states 
undergo pure rotations around some spherical triangle on the 
Bloch sphere defined by the points $A,B,C$ and $A',B',C'$, 
respectively. According to the theory of geometric phases the 
final states pick up the phases $-\frac{1}{2} \Omega (A,B,C) 
\equiv -\frac{1}{2} \Omega$ and $-\frac{1}{2} \Omega (A',B',C') 
\equiv -\frac{1}{2} \Omega'$, respectively. Thus, the total 
state $\Pi_{0} \longrightarrow \Pi_{f} = \Pi_{0}$ acquires 
the relative phase \cite{brendel95}
\begin{equation} 
\phi = \arg \langle \Pi_{0} | \Pi_{f}\rangle = 
-\frac{\Omega + \Omega'}{2} , 
\label{eq:productphase}
\end{equation}
after traversing the pair of loops. 

Now, in presence of entanglement this additive property has 
to be modified as follows. Any entangled two-photon polarisation 
state can be written on Schmidt form as 
\begin{equation}
|\Pi_{0} \rangle = \sqrt{\lambda} |AA' \rangle + \sqrt{1-\lambda} 
|A_{\perp} A_{\perp}' \rangle , 
\end{equation}
where $\langle A|A_{\perp} \rangle = \langle A'|A_{\perp}' 
\rangle =0$. The real-valued parameter $0 \leq \lambda \leq 1$ 
determines the degree of entanglement $\Xi \equiv |1-2\lambda|$ 
with $\Xi$ ranging from $0$ (maximally entangled states) to 
$1$ (product states). Suppose again that each of the two 
states $A$ and $A'$ is rotated around some spherical triangle 
so that they pick up the phases $-\frac{1}{2} \Omega$ and 
$-\frac{1}{2} \Omega'$, respectively. On the other hand, the 
orthogonal states $A_{\perp}$ and $A_{\perp}'$ are taken 
around paths with opposite orientation, but enclose the same 
area so that they pick up the phases $+\frac{1}{2} \Omega$ 
and $+\frac{1}{2} \Omega'$, respectively. Thus, the photons 
acquire the relative phase 
\begin{equation} 
\tilde{\phi} = \arg \langle \Pi_{0} | \Pi_{f}\rangle = 
\arctan \Big[ (1-2\lambda) \tan \frac{\Omega + \Omega'}{2} \Big] , 
\label{eq:entangledphase}
\end{equation}
and the visibility 
\begin{equation} 
| \langle \Pi_{0} | \Pi_{f}\rangle | = 
\sqrt{\cos^{2} \frac{\Omega + \Omega'}{2} + 
(1-2\lambda)^{2} \sin^{2} \frac{\Omega + \Omega'}{2}} \leq 1 . 
\label{eq:entangledvisibility}
\end{equation}
is reduced as the state now is noncyclic, i.e. $\Pi_{0} 
\longrightarrow \Pi_{f} \neq \Pi_{0}$. Comparing Eqs. 
(\ref{eq:productphase}) and (\ref{eq:entangledphase}) 
we may write 
\begin{equation}
\left| \frac{\tan \tilde{\phi}}{\tan \phi} \right| = \Xi , 
\label{eq:nonlinear}
\end{equation}
which demonstrates that the degree of entanglement is the 
cause of the nonlinear relation between $\tilde{\phi}$ and 
the one-photon solid angles $\Omega$ and $\Omega'$. This 
nonlinearity is particularly striking in the maximally 
entangled case, where the right-hand side of Eq. 
(\ref{eq:nonlinear}) vanishes and $\tilde{\phi}$ 
can only take the values $0$ or $\pi$. 
 
An interesting relation between the two-photon phase and the 
mixed state phase discussed in the previous section may be 
obtained by letting the loop $A',B',C'$, say, be reduced to 
a point. In this case, the relative two-photon phase becomes 
\begin{equation} 
\phi = \arg \langle \Pi_{0} | \Pi_{f}\rangle = 
\arctan \Big[ (1-2\lambda) \tan \frac{\Omega}{2} \Big] ,  
\end{equation}
which is the mixed state phase in Eq. (\ref{eq:mixedsolid}) if 
we put $\lambda = (1+r)/2$. This illustrates the fact that 
a mixed state can always be lifted into a pure state by 
attaching an ancilla. Thus, the mixed state phase for a
single polarised photon is equivalent to that obtained 
for a pure polarisation entangled two-photon state by keeping 
the polarisation of the ancilla photon fixed.  

An experiment to investigate the interference between 
$|\Pi_{0} \rangle$ and $|\Pi_{f} \rangle$ was proposed 
in \cite{hessmo00}. This experiment is based on sending  
a polarisation entangled photon pair through a Franson 
type interferometer \cite{franson89}. In the longer 
arms local rotations are applied in such a way that 
$|\Pi_{f} \rangle$ is obtained. In one of the shorter 
arms the $U(1)$ shift $e^{i\chi}$ is added to $|\Pi_{0} \rangle$.
To observe a superposition of $|\Pi_{0} \rangle$ 
and $|\Pi_{f} \rangle$ we require that the source produces 
photon pairs randomly \cite{franson89}. This is the 
case with the source for polarisation entangled photon 
pairs demonstrated experimentally in \cite{white99}. 
If the photons arrive simultaneously in the output 
detectors, they both either took the shorter path 
($\Pi_{0}$) or the longer path ($\Pi_{f}$). Thus, the  
measured coincidence profile becomes 
\begin{equation}
I = \Big| e^{i\chi}|\Pi_{0} \rangle + 
|\Pi_{f} \rangle \Big|^{2} = 
2 + 2 |\langle \Pi_{0}|\Pi_{f} \rangle| 
\cos [\chi - \arg \langle \Pi_{0}|\Pi_{f} \rangle ] ,
\end{equation}
which verifies the two-photon Pancharatnam phase in Eq. 
(\ref{eq:entangledphase}) and the reduced visibility in 
Eq. (\ref{eq:entangledvisibility}).  

\section{Test of geometric phase} 
Consider the continuous path $\eta : t \in [0,\tau ] \longrightarrow 
|A_{t} \rangle \langle A_{t}| = U_{t} |A_{0} \rangle \langle A_{0}|
U_{0}^{\dagger}$ of normalised pure state projectors with 
$\langle A_{0} | A_{\tau} \rangle \neq 0$ and $U_{t}$ unitary. 
Dividing the path into small pieces and using the additive 
property a) described in Sec. {\bf 1}, we obtain the geometric 
phase associated with $\eta$ as 
\begin{equation} 
\phi_{g} [\eta] = 
\lim_{N\longrightarrow \infty} \arg \Big( 
\langle A_{0} | A_{\tau} \rangle \  
\langle A_{\tau} | A_{[(N-1)\tau /N]} \rangle \ ... \  
\langle A_{[\tau /N]} | A_{0} \rangle \Big) . 
\label{eq:puregp} 
\end{equation} 
$\phi_{g} [\eta]$ is a property of the path $\eta$ as it is 
independent of the lift $\eta \longrightarrow \tilde{\eta} : 
t \in [0,\tau ] \longrightarrow |A_{t} \rangle$ and, in the 
$N\longrightarrow \infty$ limit, of the particular subdivision 
of the path. A parallel lift is defined by requiring that 
each $\langle A_{[(j+1)\tau /N]} | A_{[j\tau /N]} \rangle$, 
$j=0,...,N-1$, be real and positive, which in the 
$N\longrightarrow \infty$ limit takes the form 
\begin{equation}
\langle A_{t} | \dot{A}_{t} \rangle = 0 .
\label{eq:pc} 
\end{equation} 
For such a parallel lift, we obtain $\phi_{g} [\eta]$ as 
\begin{equation}
\phi_{g} [\eta] = \arg \langle A_{0} | A_{\tau} \rangle . 
\label{eq:pureparallel} 
\end{equation} 

We here argue that the geometric phase for any such path $\eta$ 
can be made experimentally accessible as a Pancharatnam 
relative phase. The idea is based on Uhlmann's \cite{uhlmann86} 
approach to the geometric phase for any quantal state 
$\rho_{t}$ using the purification $W_{t}$ such that 
$W_{t} W^{\dagger}_{t} = \rho_{t}$. For the unitary pure 
state path $\eta$ we may write $W_{t} = U_{t} |A_{0} \rangle 
\langle A_{0}| \tilde{U}_{t}$, where an ancilla state 
$|\tilde{A}_{t} \rangle = \tilde{U}_{t} |A_{0}\rangle$ is 
attached to the system. Now, a parallel purification is 
obtained by choosing $\tilde{U}_{t}$ so that  
\begin{equation}
W^{\dagger}_{t} \dot{W}_{t} = {\mbox{hermitian}} . 
\label{eq:uhlmannpc} 
\end{equation} 
For such a parallel purification, the quantity   
\begin{equation} 
\beta = \arg {\mbox{Tr}} [W_{0}^{\dagger} W_{\tau}] , 
\label{eq:uhlmannholonomy1} 
\end{equation} 
takes, in the pure state case, the form 
\begin{equation} 
\beta = \arg \langle \tilde{A}_{\tau} | A_{\tau} \rangle =  
\arg \langle A_{0} |\tilde{U}_{\tau}^{\dagger} U_{\tau} 
| A_{0} \rangle ,  
\label{eq:uhlmannholonomy2} 
\end{equation} 
which can be shown \cite{uhlmann93} to fulfil $\beta = \phi_{g} [\eta]$.  

Now, for pure states the purification does not need any 
entanglement between the system and the ancilla. Thus, the above 
procedure should work as a superposition effect for the system 
alone. To see this, consider the lift $\eta \longrightarrow 
|A_{t} \rangle = U_{t} |A_{0}\rangle$ and the auxiliary path 
$|\tilde{A}_{t} \rangle = \tilde{U}_{t} |A_{0}\rangle$, both 
in the Hilbert space of the system. By letting the auxiliary 
path be exposed to a $U(1)$ shift $e^{i\chi}$, the interference 
profile at each instant of time $\tau$ reads 
\begin{equation}
I = \Big| e^{i\chi}|\tilde{A}_{\tau} \rangle + |A_{\tau}\rangle 
\Big|^{2} = 2 + 2 |\langle \tilde{A}_{\tau}|A_{\tau} \rangle| 
\cos [ \chi - \arg \langle \tilde{A}_{\tau}|A_{\tau} \rangle ]
\end{equation}  
that attains its maximum at the relative phase 
$\phi_{\tau} = \arg \langle A_{0}| \tilde{U}_{\tau}^{\dagger} 
U_{\tau}) |A_{0} \rangle$. This phase is generally related to 
the geometric phase as $\phi_{\tau} = \phi_{g} [\eta] + 
\gamma_{\tau}$, where $\gamma_{\tau}$ is the accumulation 
of local phase changes along $\eta$. However, in analogy 
with the above purification approach, $\gamma_{\tau}$ can 
be made to vanish at each instant of time by choosing the 
auxiliary $\tilde{U}_{t}$ so that it fulfils the parallelity 
condition (\ref{eq:uhlmannpc}). In other words, if $\tilde{U}_{t}$ 
fulfils Eq. (\ref{eq:uhlmannpc}) it is assured that $\phi_{\tau}$ 
is identical to the geometric phase associated with the 
continuous path $\eta$. 

To illustrate this idea, consider a spin$-\frac{1}{2}$ initially 
polarised in the $+z$ direction and undergoing precession 
around the axis ${\bf n} = (\sin \theta ,0,\cos \theta )$ 
under the action of the Hamiltonian   
\begin{equation}
H = \frac{1}{2} \Big( \sin \theta \sigma_{x} + 
\cos \theta \sigma_{z} \Big) ,  
\end{equation}
$\sigma_{x}$ and $\sigma_{z}$ being the standard Pauli spin 
operators in the $x$ and $z$ direction, respectively. If we 
put $\tilde{U}_{t} = \exp (-i\tilde{H}t)$ and insert 
$|+z \rangle \langle +z|$ and $H$ into Eq. (\ref{eq:uhlmannpc}), 
we obtain  
\begin{equation}
\tilde{H} = \frac{1}{2} \cos \theta \sigma_{z} .   
\end{equation} 
Thus, the Pancharatnam phase difference becomes 
\begin{equation}
\phi_{\tau} = 
\langle +z | \tilde{U}_{\tau}^{\dagger} U_{\tau} |+z \rangle = 
- \arctan \Big( \cos \theta \tan \frac{\varphi}{2} \Big) +  
\frac{\varphi}{2} \cos \theta    
\label{eq:purenoncyclicgp}
\end{equation}
with $\varphi$ the precession angle of the spin. This is precisely 
the noncyclic geometric phase $-\frac{1}{2} \Omega_{gc}$,
where $\Omega_{gc}$ is the solid angle enclosed by the path 
$\eta$ and the shortest geodesic connecting its end-points on 
the Bloch sphere \cite{mukunda93}. 

Furthermore, the state $|-z\rangle \langle -z|$ acquires 
the relative phase  $+\frac{1}{2} \Omega_{gc}$ under the 
action of the unitarities $U_{t}$ and $\tilde{U}_{t}$. Thus, 
according to Eq. (\ref{eq:mixedsolid}) we obtain 
\begin{equation}
\Phi_{\tau} = - \arctan \Big[ r \tan \frac{\Omega_{gc}}{2} \Big] , 
\label{eq:mixednoncyclicgp}
\end{equation}
which is the mixed state generalisation of Eq. 
(\ref{eq:purenoncyclicgp}). An explicit neutron interferometer 
experiment implementing the procedure described in this section 
has been proposed in \cite{sjoqvist01a}. This experiment tests 
the geodesically closed solid angle in noncyclic precession of 
the neutron spin.  

\section{Dual set up}
It has been demonstrated \cite{wagh98} that Pancharatnam's 
relative phase for an internal spin degree of freedom may be 
tested in interferometry. In such an experiment the spin state 
is changed due to some local interaction in one of the beams, 
while the other beam is exposed to a variable $U(1)$ shift. Here, 
we argue that the spatial beam-pair itself may acquire an unambiguous 
Pancharatnam relative phase, dual to the usual phase acquired by 
the spin state. This dual phase could be tested polarimetrically 
by letting both a nontrivial spin-space interaction and a spin 
measurement be delocalised to both the spatial beams. 

Let us first briefly review the Pancharatnam relative phase 
for a precessing spin$-\frac{1}{2}$ system. The evolution of 
the state vector reads  
\begin{eqnarray}
|A_{0} \rangle & = &  
\cos \frac{\theta}{2} |+z\rangle + 
\sin \frac{\theta}{2} |-z\rangle \longrightarrow 
\nonumber \\ 
|A_{f} \rangle & = & 
e^{-i\varphi /2} \cos \frac{\theta}{2} |+z\rangle + 
e^{i\varphi /2} \sin \frac{\theta}{2} |-z\rangle , 
\label{eq:spinevolution}
\end{eqnarray}
where $\varphi$ is the precession angle and $|\pm z \rangle$ 
are eigenvectors of the Pauli spin operator $\sigma_{z}$ in the 
$z$ direction. The Pancharatnam relative phase between 
$|A_{0} \rangle$ and $|A_{f} \rangle$ is 
\begin{equation} 
\phi = \arg \langle A_{0} | A_{f} \rangle = 
-\arctan \Big( \cos \theta \tan \frac{\varphi}{2} \Big) , 
\label{eq:spinp}
\end{equation} 
where we have assumed that $\langle A_{0} | A_{f} \rangle 
\neq 0$. This phase could be tested in neutron interferometry 
by letting one beam be exposed to a time-independent uniform 
magnetic field in the $z$ direction and applying a variable 
phase shift $e^{i\chi}$ to the other beam. The last beam-splitter 
in the interferometer makes the two spin states $A_{0}$ and 
$A_{f}$ to interfere, yielding the output interference profile as 
\cite{wagh98}
\begin{equation} 
I = \Big| e^{i\chi} |A_{0} \rangle + 
|A_{f} \rangle \Big|^{2} =  
2+2|\langle A_{0} | A_{f} \rangle| 
\cos [\chi - \arg \langle A_{0} | A_{f} \rangle]   
\label{eq:spinintensity}
\end{equation}
with visibility 
\begin{equation}
|\langle A_{0} | A_{f} \rangle| = 
\sqrt{1-\sin^{2} \theta \sin^{2} \frac{\varphi}{2}} . 
\label{eq:spinv} 
\end{equation}

Now, the basic set up to test the dual Pancharatnam relative phase 
was proposed in \cite{sjoqvist01b} for neutrons and is sketched in 
Fig. 1. It consists of three steps: 
\begin{itemize}

\item[(i)] A superposed spatial beam state is created by 
sending a beam of neutrons polarised along the $+z$ direction 
through a beam-splitting crystal plate with transmission 
probability $T$ and reflection probability $R$. With $0$ 
and $1$ the two spatial beam states, this yields the total 
state vector  
\begin{equation}
|\Psi_{0} \rangle = |A_{0} \rangle |+z \rangle = 
\Big[ \cos (\theta /2) |0\rangle + 
\sin (\theta /2) |1\rangle \Big] |+z\rangle , 
\label{eq:outplate}
\end{equation}
where we have set $\sqrt{T} = \cos (\theta /2)$. 

\item[(ii)] The $0-$beam is exposed to a magnetic field 
${\bf B}_{0} = B_{0} \hat{{\bf x}}$ and the $1-$beam is 
exposed to ${\bf B}_{1} = B_{1} \hat{{\bf x}}$, as shown 
in Fig. 1. ${\bf B}_{0}$ and ${\bf B}_{1}$ are time-independent 
and uniform over some finite spatial region. The change in the 
spin state in each beam is described by the unitary operators 
\begin{eqnarray}
U ( \varphi_{0} ) & = & e^{-i\varphi_{0} \sigma_{x} /2} , 
\nonumber \\ 
U ( \varphi_{1} ) & = & e^{-i\varphi_{1} \sigma_{x} /2} , 
\end{eqnarray}
where $\sigma_{x}$ is the Pauli spin operator in the $x$ 
direction. Here, $\varphi_{0} \propto B_{0}$ and 
$\varphi_{1} \propto B_{1}$ are the Larmor 
precession angles. By introducing the spatial state vectors 
\begin{eqnarray}
|A^{+} \rangle & = & 
e^{- i\Delta \varphi /4} \cos \frac{\theta}{2} |0 \rangle + 
e^{i\Delta \varphi /4} \sin \frac{\theta}{2} |1 \rangle , 
\nonumber \\ 
|A^{-} \rangle & = & 
e^{i\Delta \varphi /4} \cos \frac{\theta}{2} |0 \rangle + 
e^{- i\Delta \varphi /4} \sin \frac{\theta}{2} |1 \rangle  
\label{eq:finalspatial}
\end{eqnarray}
with $\Delta \varphi = \varphi_{0} - \varphi_{1}$ and   
$\chi = (\varphi_{0} + \varphi_{1} )/2$, we may write the 
final total state vector as 
\begin{eqnarray}
|\Psi_{f} \rangle & = & \frac{1}{2} 
\Big[ e^{-i\chi /2} |A^{+} \rangle + 
e^{i\chi /2} |A^{-} \rangle \Big] |+z\rangle
\nonumber \\ 
 & & + \frac{1}{2} 
\Big[ e^{-i\chi /2} |A^{+} \rangle - 
e^{i\chi /2} |A^{-} \rangle \Big] |-z\rangle . 
\end{eqnarray}

\item[(iii)] A spin analyser measures the spin in the $z$ 
direction in each beam, as shown in Fig. 1. The interference  
profile $I$ in the $+z$ spin channel, say, is proportional to 
the sum of $+z$ detections in the two spatially separated 
analysers. This is dual to the interference measurement in a 
single spatial channel, while summing over the two spin states, 
in the set up designed to measure the spin$-\frac{1}{2}$ 
Pancharatnam relative phase \cite{wagh98}. The interference 
profile in the $+z$ spin channel reads   
\begin{eqnarray}
I & = & \Big| e^{i\chi /2} |A^{-} 
\rangle + e^{-i\chi /2} |A^{+} \rangle 
\Big|^{2} \nonumber \\ 
 & \propto & 2+2 | \langle A^{-} | A^{+} \rangle | 
\cos \Big[ \chi - \arg \langle A^{-} | A^{+} \rangle \Big] . 
\label{eq:spatint}
\end{eqnarray}
Thus, by keeping $\Delta \varphi \propto B_{0} - B_{1}$ 
fixed and varying $\chi \propto B_{0} + B_{1}$ the dual 
Pancharatnam relative phase $\arg \langle A^{-} | A^{+} \rangle$ 
and visibility $| \langle A^{-} | A^{+} \rangle |$ can be 
measured. 

\end{itemize}

\section{Conclusions}
We have discussed extensions of Pancharatnam's prescription 
of relative phase to mixed states and entangled two-photon 
states. The solid angle relation for single qubit states 
becomes nonlinear in these extensions and we have discussed 
experiments to test this nonlinearity. A continuously evolving 
quantal state is associated with a geometric phase at each 
point along the path in state space. For a given lift of such 
a path we have proposed a general procedure to systematically 
cancel the accumulation of local phase changes relative a 
specific auxiliary path in Hilbert space. This guarantees 
that Pancharatnam's relative phase between the two paths 
in Hilbert space is identical to the geometric phase. Finally, 
we have argued that a spatially delocalised state may acquire 
a dual Pancharatnam relative phase that could be measured 
polarimetrically. We hope that these ideas may lead to new 
experimental tests of Pancharatnam's prescription of relative 
phase as well as to trigger further extensions to the theory 
of interference.

\section*{Acknowledgements} 
I wish to thank Jeeva Anandan, Artur Ekert, Marie Ericsson, 
Bj\"{o}rn Hessmo, Daniel Oi, Arun Pati, and Vlatko Vedral 
for collaboration on Pancharatnam's phase. This work was 
supported by the Swedish Research Council.

\newpage 
\section*{Figure Captions} 
Fig.1. Set up for testing the dual Pancharatnam relative 
phase. The source prepares a coherent superposition of the 
two spatial beam states $0$ and $1$, 
both being spin-polarised perpendicular to the plane of the 
set up. Each beam is exposed to a magnetic field in a common 
direction in the plane of the set up, but with different strength 
$B_{0} \neq B_{1}$. The dual Pancharatnam relative phase is 
measured by the pair of spin analyzers by keeping the difference 
$B_{0} - B_{1}$ fixed, while varying the sum $B_{0} + B_{1}$.
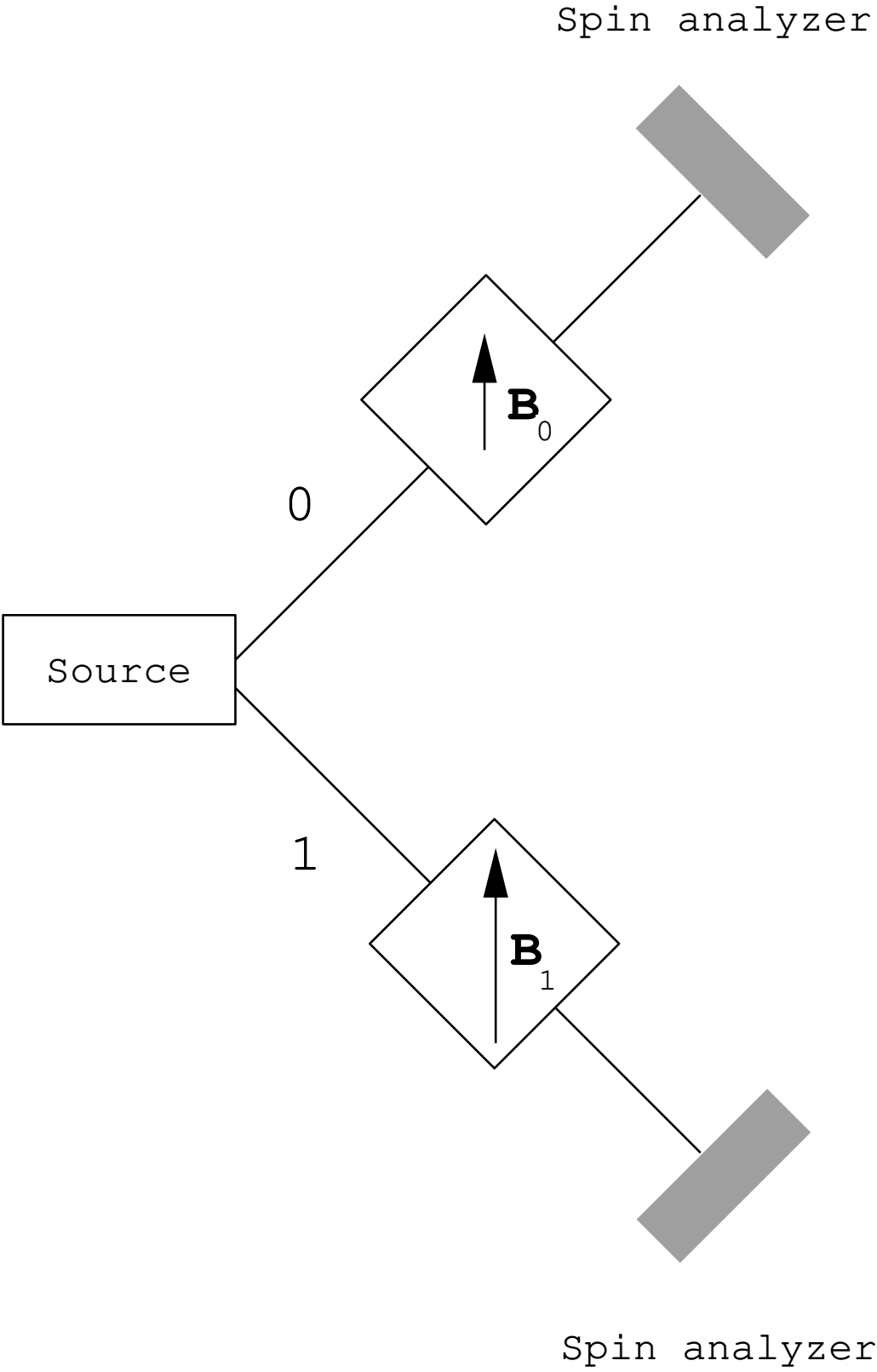
\end{document}